\documentclass[11pt]{article}
\usepackage{graphicx}

\newcommand{\h}{\hspace*{5 ex}}

\newcommand{\noi}{\noindent}
\begin{document}

\Large{\bf {ETTORE MAJORANA:}}\\
\Large{\bf {HIS WORK AND HIS LIFE}}

\

April 2014 \hfill\break

\

Erasmo Recami

\vspace*{0.1 cm}

{{\em Facolt\`{a} di Ingegneria, Universit\`{a} statale di Bergamo,
Dalmine (BG), Italy;}}

{{\em INFN---Sezione di Milano, Milan, Italy; {\rm and}}}

{{\em DMO/FEEC, UNICAMP, Campinas, SP, Brazil}}

\baselineskip 0.4cm

\

\

{\bf Abstract --- } \ In this paper we present a panoramic view of the main scientific articles published by Ettore Majorana, the brightest Italian theoretical physicist of the XX century (actually, Enrico Fermi regarded him as the brighest in the world of his time, and compared him to Galileo and Newton; even if to some people Majorana is often known mainly for his mysterious disappearance in 1938, when he was 31). Extensive information and comments are added with regard to the scientific manuscripts left unpublished by him. We also outline his life, the biographical data being based on letters, documents, testimonies discovered or collected by the author during more than four decades, and contained for instance in Recami's 1987 book quoted in the text. Two pictures complete the paper.

\

\section{Historical touch}

Ettore Majorana's fame\cite{ER}  solidly rests on
testimonies like the following one, by the mindful pen of Giuseppe Cocconi.
At the request of Edoardo Amaldi\cite{Amaldi}, he wrote from CERN (July $18^{\rm th}$, 1965):

\vskip2mm
\h     ``In January 1938, after having just graduated, I was offered,
essentially by you, to come to the Institute of Physics at the
University in Rome for six months as a teacher assistant, and once I
was there I would have the good fortune of joining Fermi, Bernardini
(who had been granted a chair at Camerino a few months earlier) and Ageno
(he too, a neo-graduate), in the research of
the products of disintegration of ``mesons" $\mu$ (at that time called
mesotrons or yukons) which are produced by cosmic rays. [...]

\h     It was actually while I was staying with Fermi in the little
laboratory on the second floor, absorbed in our work, with Fermi working
a piece of Wilson's chamber (which would help to reveal mesons at the
end of their range) on a lathe and me constructing a jalopy for the
illumination of the chamber, using the flash produced by the explosion
of an aluminum ribbon shortcircuited on a battery, that Ettore Majorana
came in search of Fermi.  I was introduced to him and we exchanged few
words.  A dark face.  And that was it.  An easily forgettable
experience if, after a few weeks while I was still with Fermi in that
same workshop, news of Ettore Majorana's disappearance in Naples had not
arrived.  I remember that Fermi busied himself with telephoning around
until, after some days, he had the impression that Ettore would never be
found.

\h It was then that Fermi, trying to make me understand the
significance of this loss, expressed himself in quite a peculiar way:
he, who was so objectively stern when judging people.  And so at this
point I would like to repeat his words, just as I can still hear them
ringing in my memory: `Because, you see, in the world there are various
categories of scientists; people of a secondary or tertiary standing,
who do their best but do not go very far.  There are also those of high
standing, who come to discoveries of great importance, fundamental for
the development of science' (and here I had the impression that he had
placed himself in that category).  `But then there are geniuses like
Galileo and Newton.  Well, Ettore was one of them.  Majorana had what no
one else in the world had [...]'."

\h     And with first-hand knowledge, Bruno Pontecorvo, added: ``some time
after his entry into Fermi's group, Majorana already possessed such an
erudition and had reached such a high level of comprehension of physics
that he was able to speak on the same level with Fermi about scientific
problems.  Fermi himself held him to be the greatest theoretical
physicist of our time.  He often was astounded [...].  I remember
exactly these words that Fermi spoke: `If a problem has already been
proposed, no one in the world can resolve it better than Majorana'." (see
\cite{Ponte,ER1,ER2}.)

\h  Ettore Majorana disappeared rather mysteriously on
March 26$^{\rm th}$, 1938, and nobody
found him \cite{ER,ER1,ER2}. The myth of his ``disappearance"  has contributed
to nothing more than the notoriety he was entitled to, for being a true
genius and a genius before his time. \ All the serious documents known till
now have been discovered or collected by us (starting with 1970), and published
for the first time by us, for instance in our book \cite{ER}.

 \h    In more recent times, after that his lecture notes had been published\cite{Lezioni}, we eventually
published his notebooks {\em Volumetti}, namely, his study notes written in Rome
between 1927, when he moved from his studies in engineering to take up
physics, and 1931. Those manuscripts are a model not only of order, divided
by argument and even supplied with an index, but also in conciseness,
essentiality and originality. So much so that the related notebooks
can be regarded in a sense as a modern text of
theoretical physics, even after ninty years, and a ``mine" of new theoretical,
physical, mathematical ideas and hints, quite stimulating and useful for
present time research\cite{Kluwer}. \ Even more interesting for present time physics are
the {\em pure research} scientific manuscripts (like his {\em Quaderni}) left unpublished by Majorana: we
made recently known part of them in a 500 pages Springer's volume\cite{Springer}. \ Incidentally, catalogues of
the scientific manuscripts left by Majorana can be found in \cite{catalog,Springer}.

\h     Let us recall\cite{ER} that Majorana, after having switched to
physics at the beginning of 1928, graduated with Fermi on July 6$^{\rm th}$, 1929,
and went on to cooperate with the famous group created by Enrico Fermi
and Franco Rasetti (at the beginning, with Orso M. Corbino's important help); whose
theory subgroup was formed mainly, in the order of their
entrance into the Institute, by Ettore Majorana, Gian Carlo
Wick, Giulio Racah, Giovanni Gentile Jr., Ugo Fano,
Bruno Ferretti, and Piero Caldirola; while the members of the
experimental subgroup were: Emilio Segr\'e, Edoardo
Amaldi, Bruno Pontecorvo, Eugenio Fubini, Mario Ageno,
Giuseppe Cocconi, as well as the chemist Oscar D'Agostino. \
Afterwards, Majorana qualified for university
teaching of theoretical physics (``Libera
Docenza") on November 12$^{\rm th}$, 1932; spent about six months in
Leipzig with W. Heisenberg during 1933; and then for some
unknown reasons stopped partecipating in the Fermi's group
activities and even in publishing his works. Except for his
paper ``Teoria simmetrica dell'elettrone e del positrone"
that (even if ready since 1933) Majorana was happily convinced by his
colleagues to take out from a drawer and publish at the
threshold of the 1937 Italian national competition for three
full-professorships: That is the paper dealing with the ``Majorana's neutrino".

\h   With respect to the last point, let us recall that in 1937
there were numerous Italian competitors and many of them were of
exceptional valor; above all: Ettore Majorana,
Giulio Racah, Gian Carlo Wick, and Giovanni Gentile Jr. (the
son of the famous philosopher bearing the same name, and the inventor of
``parastatistics" in quantum mechanics). The judging committee
was presided by E.Fermi and had as members E. Persico,
G. Polvani, A. Carrelli and O. Lazzarino. After a proposal by the
judging commettee, the Italian Minister of National
Education nominated Majorana professor of theoretical physics
at Naples University because of his ``great and well-deserved fame",
independently of the competition itself (actually,
``the Commission hesitated to apply the normal university
competition procedures to him"). The attached report
on the scientific activities of Ettore Majorana, sent to the minister by
the committee, states\cite{ER}:

\h    ``Without listing his
works, all of which are highly notable both for their originality of
methods utilized as well as for the importance of the achieved results,
we limit ourselves to the following:

\h    In modern nuclear theories the contribution made by this researcher
with the introduction of the forces called ``Forces of Majorana" is
universally recognized as, among the most fundamental, the one that
permits us to theoretically comprehend the reasons for nuclear
stability.  The work of Majorana today serves as a base for the most
important research in the field.

\h    In atomic physics, the merit of having resolved some of the most
intricate questions on the structure of spectra through simple and
elegant considerations of symmetry is due to Majorana.

\h    Lastly, he devised a brilliant method that permits us to treat the
positive and negative electron in a symmetrical way, finally eliminating
the necessity of relying on the extremely artificial and unsatisfying
hypothesis of an infinitely large electrical charge diffused in the
space, a question that had been approached in vain by many other
scholars."

One of the most important works of Ettore Majorana, the one that
introduces his ``infinite components equation" was not mentioned: it
had not been understood yet.  It is
interesting to note, however, that the right light was shed on his theory
of electron and anti-electron symmetry (today at its height for its
application to neutrinos and anti-neutrinos, as we mentioned); and in result of his
capacity to eliminate the hypothesis called ``of the Dirac sea; a hypothesis
that was defined as ``extremely artificial and unsatisfying", regardless
of the fact that in general it had been uncritically accepted.

\h    The details of when Majorana and Fermi met the first time were
narrated by E. Segr\'e \cite{ES}: ``The first important work written by Fermi
in Rome [{\it Su alcune propriet\`a statistiche dell'atomo} (On certain
statistical properties of the atom)] is today known as the Thomas-Fermi
method... When Fermi found that he needed the solution to a non-linear
differential equation characterized by unusual boundary conditions in order
to proceed, in a week of assiduous work with his usual energy, he
calculated the solution with a little hand calculator.  Majorana, who
had entered the Institute just a short time earlier and who was always
very skeptical, decided that probably Fermi's numeric solution was wrong
and that it would have been better to verify it.  He went home,
transformed Fermi's original equation into a Riccati equation, and
resolved it without the aid of any calculator, utilizing his extraordinary
aptitude for numeric calculation. When he returned to the Institute and
skeptically compared the little piece of paper on which he had written his
results to Fermi's notebook, and found that the results coincided
exactly, he could not hide his amazement."
We have taken the liberty of
indulging in an anecdote since the pages in which Majorana solved that
equation have been eventually found by us, and it has been shown elsewhere that
he did actually follow two independent (and quite original) mathematical
methods to the same result: one of them leading to an Abel equation, rather
than to a Riccati equation.

\section{Ettore Majorana's published papers}

Majorana published few scientific articles: nine, actually, besides
his semi-popular work entitled ``Il valore delle leggi statistiche
nella fisica e nelle scienze sociali" (The value of statistical laws in
physics and social sciences), that was however published not by
Majorana, but (posthumously) by G.Gentile Jr., in Scentia [36 (1942) 55-56]. \
We already know that
Majorana changed over from engineering to physics in 1928 (the year in
which he published his first article, written in collaboration with his
friend G.Gentile jr.), and then went on publishing his works in theoretical
physics only for very few years, practically until 1933 only. \ Nevertheless,
even his {\em published} works are a goldmine of ideas and techniques of
theoretical physics that still remains partially unexplored.
Let us list his nine articles:
\begin{enumerate}
\item[{1)}]
  ``Sullo sdoppiamento dei termini Roentgen ottici a causa
dell'elettrone rotante e sulla intensit\`a delle righe del Cesio",
in coll. with Giovanni Gentile, Jr.: {\em Rendiconti Accademia
Lincei} 8 (1928) 229-233.

\item[{2)}] ``Sulla formazione dello ione molecolare di He": {\em Nuovo Cimento}
8 (1931) 22-28.

\item[{3)}] ``I presunti termini anomali dell'Elio": {\em Nuovo Cimento} 8 (1931)
78-83.

\item[{4)}] ``Reazione pseudopolare fra atomi di Idrogeno": {\em Rendiconti
Accademia Lincei} 13 (1931) 58-61.

\item[{5)}] ``Teoria dei tripletti {\em P'} incompleti": {\em Nuovo Cimento} 8
(1931) 107-113.

\item[{6)}] ``Atomi orientati in campo magnetico variabile": {\em Nuovo Cimento} 9
(1932) 43-50.

\item[{7)}] ``Teoria relativistica di particelle con momento intrinseco
arbitrario": {\em Nuovo Cimento} 9 (1932) 335-344.

\item[{8)}] ``\"Uber die Kerntheorie": {\em Zeitschrift f\"ur Physik} 82 (1933)
137-145; \ and \ ``Sulla teoria dei nuclei": {\em La Ricerca Scientifica}
4(1) (1933) 559-565.

\item[{9)}] ``Teoria simmetrica dell'elettrone e del positrone": {\em Nuovo Cimento}
14 (1937) 171-184.
\end{enumerate}

\

\noindent The first ones, written
between 1928 and 1931, regard atomic and molecular physics: mainly
questions of atomic spectroscopy or chemical bonds (within quantum mechanics,
of course).  As E. Amaldi has written \cite{Amaldi}, a profound examination of
these works leaves one struck by their high class: they reveal both a
deep knowledge of the experimental data even in the most minute
details, and an uncommon ease, above all at that time, in the use
of the symmetry properties of the quantum states in order to
qualitatively simplify the problems and choose the most opportune way
for quantitative resolution.  As to these first articles, we confine
ourselves to  quote Arimondo et al.'s paper, which associated to them
the birth, e.g., of {\em autoionization\/}.\cite{Arimondo}

\h One cannot forget also the 6th one,
``Atomi orientati in campo magnetico variabile" (Atoms oriented in
a variable magnetic field).
It is the article, famous amongst atomic physicists, in which the effect
now known as the Majorana-Brossel Effect
was introduced.  In it, Majorana predicts and calculates the modification
of the spectral line shape due to an oscillating magnetic field.  This
work has also remained a classic of the treatment of non-adiabatic
spin-flip.  His results ---once generalized, as suggested by Majorana
himself, by Rabi in 1937 and by Bloch and Rabi in 1945--- established the
theoretical base of the experimental method used to reverse the spin also
of neutrons by a radiofrequency field: a method that is still used today,
for example, in all polarized-neutron spectrometers. This article introduces
moreover the so-called ``Majorana Sphere" (to represent spinors by a set of
points on the surface of a sphere), as it has been enthusiastically noticed,
not long ago, by R.Penrose\cite{Penrose}.

\h     Majorana's last three articles too are all of such importance that none
of them can be set aside without comment\cite{ER}.

\h     The article ``Teoria relativistica di particelle con momento intrinseco
arbitrario"  (Relativistic theory of particles with arbitrary spin),
is a typical example of a work that is
so far ahead of its time that it is understood and evaluated in depth
only many years later. At that time it was common opinion that one could
write relativistic quantum equations only in the case of particles with zero
or half spin.  Convinced of the contrary, Majorana  ---as we know from his
manuscripts--- began constructing suitable
quantum-relativistic equations for the subsequent possible spin values
(one, three-halves, etc.); and he even found out the method for writing
down the equation for a generic spin-value. But he published nothing,
until he discovered that a single
equation could be written to represent an infinite series of cases, that
is, an entire infinite family of particles of any spin (even if at
that time the known particles could be counted on one hand).  In order to
carry his program out with these ``infinite components" equations, he
invents a technique for the representation of a group several years before
Eugene Wigner. And, what is more, Majorana gets the infinite dimensional
unitary representations of the Lorentz Group that will be re-discovered by
Wigner in his 1939 and 1948 works. The whole theory was re-invented by
Soviet mathematicians (in particular Gelfand and collaborators) in a series
of articles from 1948 to 1958, and finally applied by physicists in even
later years.  Majorana's initial article, actually, remained in the shadows
for a good 34 years until D.Fradkin\cite{Fradkin} released, in 1966, what Majorana had
accomplished so many years earlier.

\vskip3mm
\h As soon as the news of the Joliot-Curie experiments\cite{JC} reached Rome
at the beginning of 1932, Majorana understood that they had discovered the
``neutral proton" without having realized it.  Thus, even before the
official announcement of the discovery of the neutron, made just after
by Chadwick\cite{Ch}, Majorana was able to explain the structure and the
stability of atomic nuclei through protons and neutrons, preceding in this
way also the pioneering work of D.Ivanenko\cite{Ivanenko}: as both Segr\'e and Amaldi
have recounted.  His colleagues remember that even before Easter he had
already concluded that protons and neutrons (indistinguishable with respect
to the nuclear intercation) were bound by the ``exchange forces" originating
from the exchange of their spatial positions only (and not also of their
spins, as Heisenberg will instead propose), so as to obtain the alpha
particle (and not the deutron) as saturated in respect of the binding energy.

\h Only after that Heisenberg had published his own article on the same
argument, Fermi was able to persuade Majorana to meet the famous colleague
in Leipzig; and finally Heisenberg is able to convince Majorana to publish
his results in the paper ``Uber die Kerntheorie".
Majorana's paper on the stability of nuclei was immediately
recognized by the scientific community ---a rare event, as we know, for
his writings--- thanks to that timely ``propaganda" done by Heisenberg
himself. \ Let us seize the present opportunity for quoting two brief
passages from Majorana's letters\cite{ER} from Leipzig. On February 14, 1933,
he tells his mother: ``The environment of the physics institute is very
nice. I have good relations with Heisenberg, with Hund, and with everyone
else. {\em I am writing some articles in German.  The first one is already
ready}...". The work that is already ready is, naturally, the on nuclear
forces that we are speaking about; which, however, remained {\em the only
one} in German. Again: in the letter dated February 18$^{\rm th}$, he declares to his
father\cite{ER}: ``{\em I will publish in German, after having extended it, also my
latest article which appeared in Nuovo Cimento}."  Actually, Majorana did not
publish anything more, neither in Germany  nor after his return to Italy,
except for the article (in 1937) of which we are about to speak. It is
therefore of notable importance to know that Majorana was
writing other works: in particular, that he was expanding his article about
the infinite components equations.

\vskip3mm
\h As we have said, from the manuscripts left it appears that
Majorana was also formulating the essential lines of his symmetrical theory
of electrons and anti-electrons during those years (1932-1933).  Even though
Majorana published this theory years later, on the point of participating
in the above mentioned competition for professorship: ``Teoria simmetrica
dell'elettrone e del positrone" (Symmetrical theory of the electron and
positron); a publication that was
initially noted almost exclusively for having introduced the Majorana
representation of the Dirac matrices in a real form. A consequence of this
theory is that a neutral fermion can coincide with its anti-particle: and
Majorana suggests that neutrinos may be particles of this type. As with
Majorana's other writings, this article also started to have luck only
decades later, beginning in 1957.  Now expressions like Majorana spinors,
Majorana mass, and Majorana neutrinos are fashionable. \ As we already
mentioned, Majorana's publications (still little known, despite it all) can
be regarded as a goldmine for physics.  Recently, for example, it has been
observed by C. Becchi
how, in the first pages of this paper, a clear formulation of the
quantum action principle appears: the same principle that in following years,
for instance through Schwinger's and Symanzik's works, has brought about
quite important developments in quantum field theory.\par

\section{Ettore Majorana's unpublished papers}

Majorana also left us many unpublished scientific manuscripts (see, e.g., ref.\cite{MBR,Giannetto,SE}),
also kept at Domus Galilaeana, which have been catalogued\cite{catalog,Springer}.
The analysis of these
manuscripts has allowed us to reveal that all the existing material seems
to have been written by 1933 (even the rough copy of his last article,
that Majorana will go on to publish in 1937 ---as we already mentioned---
seems to have been ready by 1933: the year in which he had the confirmation
of the discovery of the positron).  While nothing arrived to us
of what he did in the following years from 1934 to 1938; except for a long
series of 34 letters of response, written by Majorana (between March 17$^{\rm th}$,
1931, and November 16$^{\rm th}$, 1937) to his uncle Quirino ---a renowned experimental
physicist, and a president of the Italian Physical Society--- who had been
pressing him for theoretical explanations of his own experiments.  By
contrast, his sister Maria recalled\cite{ER} that even in those years Majorana
---who had reduced his visits to the Institute starting from the beginning
of 1933 (that is, from his return from Leipzig)--- continued to study and
work at home many hours during the day and at night. Did he continue to
dedicate himself to physics?  From his letter to Quirino dated January 16,
1936, we find a first answer, because we come to learn that Majorana had been
occupied ``since some time, with quantum electrodynamics"; namely, knowing
Majorana's modesty and love for understatements, that in
1935 Majorana had profoundly devoted himself to original research in the
field of quantum electrodynamics\cite{ER}.
\vskip5mm
\h     Do any other unpublished scientific manuscripts exist?  The question
raised by his letters from Leipzig to his family becomes of greater
importance when one reads the letters sent to C.N.R.
(the National Research Council of Italy) during that period.
In the first one (dated January 21$^{\rm st}$, 1933) Majorana specifies\cite{ER}: ``At the moment,
I am occupied with the elaboration of a theory for the description of
arbitrary spin particles, that I began in Italy and of which I gave a
summary notice in Nuovo Cimento...".  In the second one (dated March 3$^{\rm rd}$,
1933) he even declares, referring to the same work: ``I have sent an article
on nuclear theory to Zeitschrift f\"ur Physik. I have the manuscript
of a new theory on elementary particles ready and will send it to the
same journal in a few days".  If we remember that the
article considered here as a ``summary notice" of a new theory was
already of a very high level, one can understand how interesting it would be
discovering a copy of the complete theory, which went unpublished.  Perhaps,
is it still in the Zeitschrift f\"ur Physik archives? \ And one must not
forget, moreover, that the above-cited letter to Quirino Majorana, dated
January 16, 1936, revealed to us that Majorana continued to work on
theoretical physics even subsequently, occupying himself in depth ---at
least--- with quantum electrodynamics.

\h     Some other of Majorana's ideas, when they did not remain in his mind,
have left a trace in his colleagues' memories. One of the testimonies we
gathered is for instance of Gian Carlo Wick.  Writing from Pisa on October
16, 1978, he says\cite{ER}:
``...The scientific contact [between Ettore and me] mentioned by
Segr\'e happened in Rome on the occasion of the ``A. Volta Congress"
(a great deal before Majorana's sojourn in Leipzig).  The conversation took
place in Heitler's company at a restaurant, and therefore without a
blackboard...; but even with the absence of details, what Majorana
described in words was a "relativistic theory of charged particles with
zero spin based on the idea of field quantization` (second quantization).
When much later I saw Pauli and Weisskopf's article [Helv. Phys. Acta
7 (1934) 709], I remained absolutely convinced that what Majorana had
discussed was the same thing...".

\h  Other papers on Majorana may be found for instance in the volume
{\em Majorana Legacy in Contemporary Physics\/}\cite{Licata}. Let us
finally recall also the papers in ref.\cite{Last}.

\

\begin{figure}[!h]
\begin{center}
 \scalebox{2.0}{\includegraphics{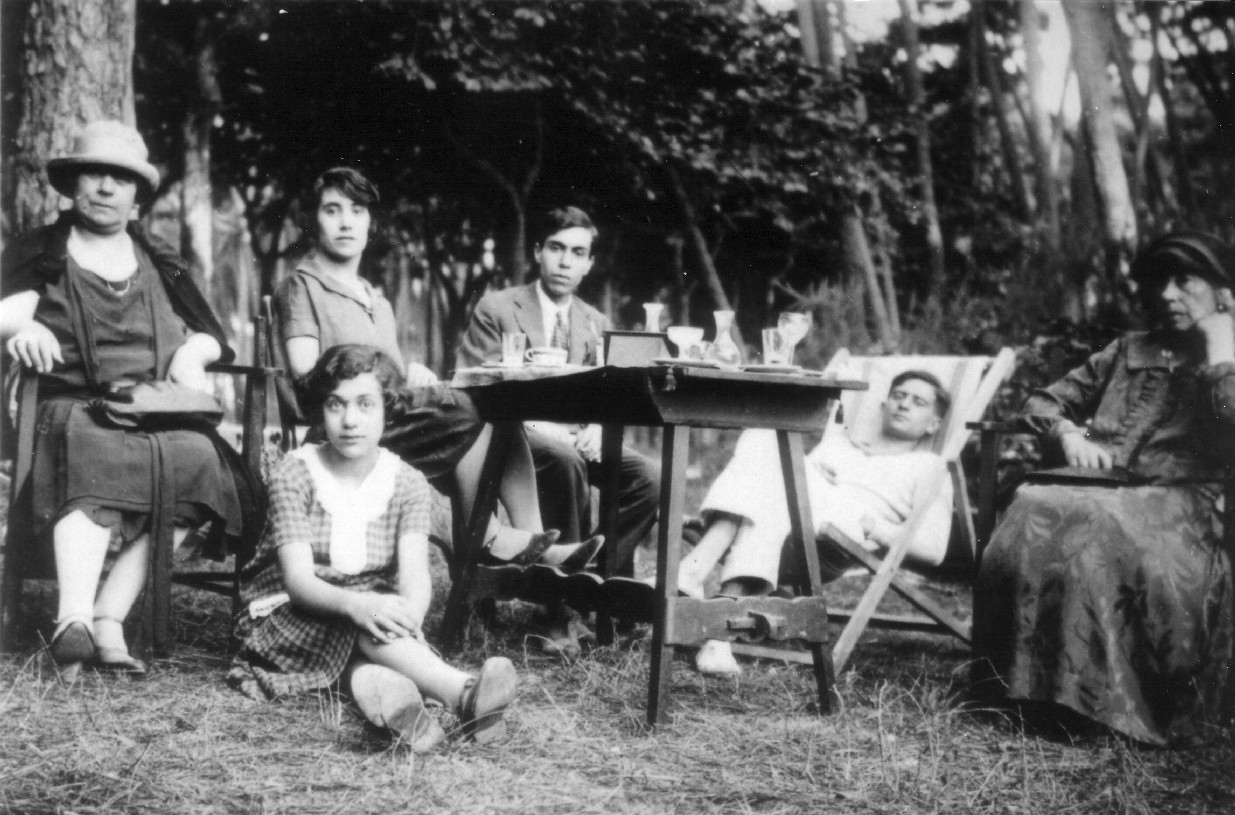}}
\end{center}
\caption{38) Ettore Majorana (at the center) in Viareggio's pinewood, Italy,
August 1926, together with --from the left-- his mother,
his sisters Maria and Rosina, his friend and fellow student
Gastone Piqu\'e, and his maternal grandmother. Courtesy of F.Majorana, B.Russo, B.Piqu\'e and
the author. {\em Reproduction forbidden}.
} \label{fig1}
\end{figure}

\

\section{Teaching theoretical physics}

As we have seen, Majorana contributed significantly to some theoretical researches
which were considered as the frontier topics in the 1930s. However, his own peculiar
contribution ranged also on the basics and the applications of quantum mechanics, and
Majorana's lectures on theoretical physics\cite{Lezioni} very effectively give evidence of this
contribution.

\h As realized only recently, Majorana revealed a genuine interest in advanced physics
teaching starting from 1933, just after that he obtained (at the end of 1932) the
professorship degree of ``libero docente'' (analogous to the German Privatdozent). Due to
this position, he proposed some academic courses at the University of Rome\cite{DeGregorio}, as testified
by the programs of three courses he would have given between 1933 and 1937. These
documents are important, since they cover a period of time that has been referred
to as Majorana's gloomy years by the testimonies of that epoch. Although Majorana never
effectively delivered such three courses, probably due to the lacking of students, they
are particularly interesting and informative due to a very careful choice of the topics
he would have treated in his courses.

\

{ \small{
The first course (academic year 1933-34) proposed by Majorana
was that of Mathematical Methods of Quantum Mechanics; the program
for it contained the following topics\cite{DeGregorio}:

\

\noi 1) Unitary geometry. Linear transformations. Hermitian
operators. Unitary transformations. Eigenvalues and eigenvectors.
2) Phase space and the quantum of action. Modifications to
classical kinematics. General framework of quantum mechanics. 3)
Hamiltonians which are invariant under a transformation group.
Transformations as complex quantities. Non compatible systems.
Representations of finite or continuous groups. 4) General
elements on abstract groups. Representation theorems. The group of
spatial rotations. Symmetric groups of permutations and other
finite groups. 5) Properties of the systems endowed with spherical
symmetry. Orbital and intrinsic momenta. Theory of the rigid
rotator. 6) Systems with identical particles. Fermi and
Bose-Einstein statistics. Symmetries of the eigenfunctions in the
center-of-mass frames. 7) The Lorentz group and the spinor
calculus. Applications to the relativistic theory of the
elementary particles.

\

\h The second course (academic year 1935-36) was instead that of
Mathematical Methods of Atomic Physics and the corresponding
arguments are\cite{DeGregorio}:

\

\noi Matrix calculus. Phase space and the correspondence principle. Minimal statistical
sets or elementary cells. Elements of the quantum dynamics. Statistical theories. General
definition of symmetry problems. Representations of groups. Complex atomic spectra.
Kinematics of the rigid body. Diatomic and polyatomic molecules. Relativistic theory of
the electron and the foundations of electrodynamics. Hyperfine structures and alternating
bands. Elements of nuclear physics.

\

\h Finally, the third course (academic year 1936-37) was that of Quantum Electrodynamics,
whose main topics were\cite{DeGregorio}:

\

\noi Relativistic theory of the electron. Quantization procedures.
Field quantities defined by commutability and anticommutability
laws. Their kinematical equivalence with sets with an undetermined
number of objects obeying the Bose-Einstein or Fermi statistics,
respectively. Dynamical equivalence. Quantization of the
Maxwell-Dirac equations. Study of the relativistic invariance. The
positive electron and the symmetry of charges. Several
applications of the theory. Radiation and scattering processes.
Creation and annihilation of opposite charges. Collisions of fast
electrons.}
}

\

\h However, Majorana effectively lectured on theoretical physics only in 1938 when, as
recalled above, he obtained a position as a full professor in Naples. He delivered his
lectures starting on January 13 and ending with his disappearance (March 26), but his
activity was intense, and his interest for teaching extremely high\cite{ER}. For the benefit of
his students, and perhaps also for writing down a book, he prepared careful notes for his
lectures. A recent analysis showed that Majorana's 1938 course was very innovative for
that time, and this has been confirmed by the retrieval (on September 2004) of a faithful
transcription of the whole set of Majorana's lecture notes (the so-called ``Moreno
lecture notes'') comprising 6 lectures not included in the original collection\cite{Lezioni}.

\h The first part of his course on theoretical physics dealt with the phenomenology of
the atomic physics and its interpretation in the framework of the old quantum theory of
Bohr-Sommerfeld. This part presents strict analogy with the course given by Fermi in Rome
(1927-28) followed by the student Majorana.

\h The second part starts, instead, with the classical radiation
theory, reporting explicit solutions of the Maxwell equations,
scattering of the solar light and some other applications. It then
continues with the Theory of Special Relativity: after the presentation of
the corresponding phenomenology, a complete discussion of the
mathematical formalism required by the theory is given, ending
with some applications as the relativistic dynamics of the
electron. Then a discussion of important effects for the
interpretation of quantum mechanics, such as the photoelectric
effect, the Thomson scattering, the Compton effects and the
Franck-Hertz experiment, follows\cite{Lezioni}.

\h The last part of the course, more mathematical in nature,
treats explicitly on quantum mechanics, both in the Schr\"odinger
and in the Heisenberg formulation. This part does not follow the
Fermi approach, but rather refers to previous personal studies by
Majorana, also following the original Weyl's book on Group Theory
and quantum mechanics.

\

\section{The ``Volumetti"}

In 2003 we have reproduced and translated for the first time, in a Kluwer's book\cite{Kluwer},
five orderly notebooks known, in Italian, as ``Volumetti" (booklets). They were
written in Rome by Ettore Majorana between 1927 and 1932.
The original manuscript is kept at the Domus Galileana in Pisa. Each of
them is composed of about 100$-$150 sequentially enumerated pages. A table of
contents is given in the first page
of each notebook. This was filled by the author when a particular section
was completed. A date, written by the author at the initial blank page of each
notebook, records when the notebook was finished.
In fact, the last book, which is the smallest one, is probably unfinished and does
not contain this information.

\h Each notebook was written during a period of time of about one year, starting
from the years in which Ettore Majorana was completing his studies at the
University of Rome. Thus the content of these notebooks goes from typical
topics covered in academic courses to frontier research ones. Despite such
a mixing between different arguments (which can be detected by looking at
different notebooks as well as in a single notebook), the style with which a topic
is treated is never obvious. As an example we refer here to the study of
the shift of the melting point of a given substance when it is placed in
a region with a magnetic field or, more interestingly, that of heat
propagation using the ``cricket simile". Also noticeable are the
contemporary physics topics treated by Ettore Majorana in an original and
very clear way, such as the Fermi explanation of the electromagnetic mass
of the electron, the Dirac equation with its applications or the Lorentz
group, revealing in some cases the preferred existing literature. As far as
frontier research arguments are concerned, we here quote only two
illuminating examples: the study of quasi-stationary states, anticipating
the Fano theory\cite{Leonardi} of about 20 years (as initially pointed out to us by the late Prof.G.F.Bassani),
and the Fermi theory of atoms,
reporting analytic solutions of the Thomas-Fermi equation with appropriate
boundary conditions in terms of simple quadratures which, to our best
knowledge, were still lacking till our recent publication (see also refs.\cite{ThomasFermi}).

\

\begin{figure}[!h]
\begin{center}
 \scalebox{2.0}{\includegraphics{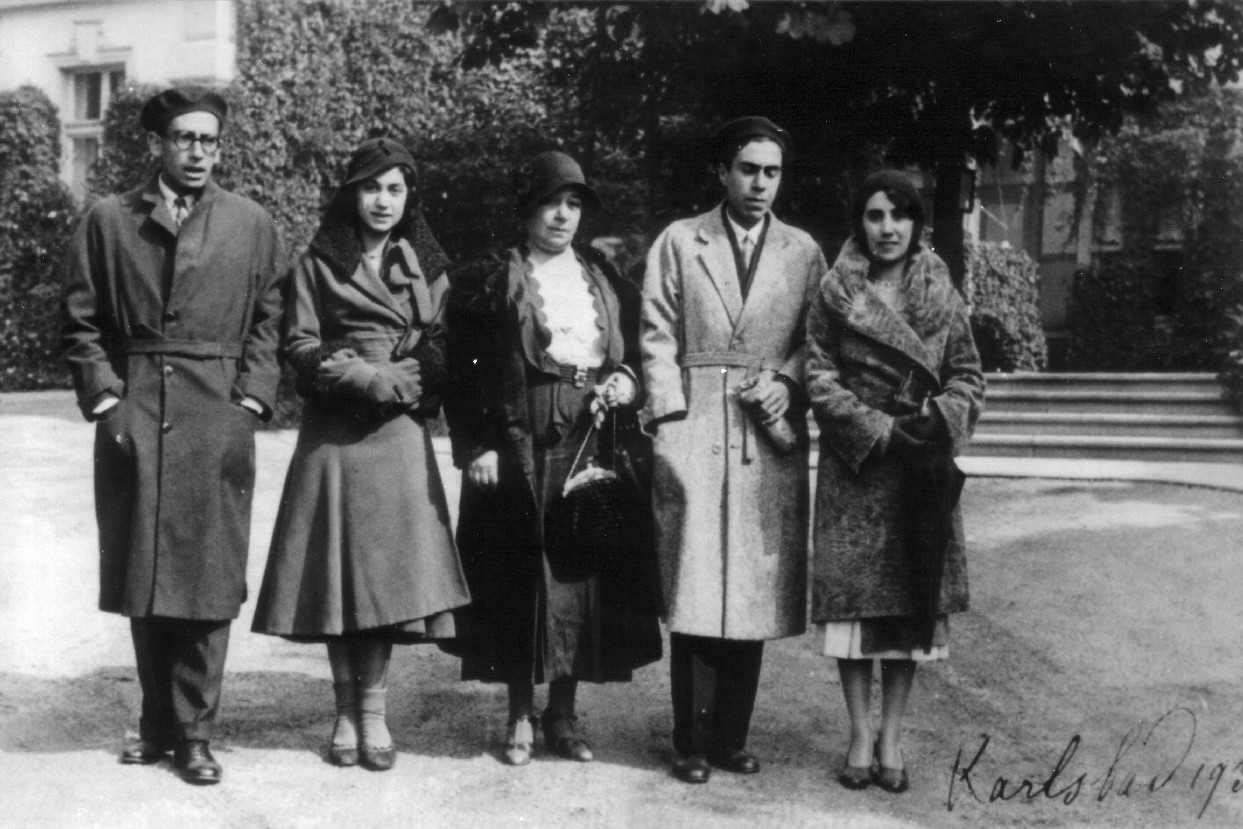}}
\end{center}
\caption{Ettore Majorana (the second from the right) together with his brother
Salvatore, his mother (at the center), and his sisters
Maria and Rosina. Karlsbad --at that time in Czechoslovakia--, autumn of 1931.
Courtesy of F.Majorana, B.Russo, B.Piqu\'e, and the author.
{\em Reproduction forbidden}.
} \label{fig2}
\end{figure}

\

\section{Majorana's research notes: the {\em ``Quaderni''}}

The material presented in the Kluwer volume, as just mentioned, is a paragon of order,
conciseness, essentiality and originality. So much so that those notebooks can be
partially regarded as an innovative text of theoretical physics, even after almost ninty
years; besides they too being a gold-mine of seminal new theoretical, physical, and mathematical
ideas and hints, stimulating and useful for modern research.

\h  But Majorana's most interesting scientific manuscripts
---namely, his {\em research} notes--- are represented by a
host of loose papers and by the {\em Quaderni}, a selection of which, as announced in advance, formed the contents
of a second book of ours\cite{Springer}, published by Springer in 2009. The topics treated in the Quaderni range from classical physics to
quantum field theory, through the study of a number of applications in atomic, molecular
and nuclear physics.

\h  A particularly relevant attention is therein reserved to the Dirac
theory and to quantum electrodynamics.

\h  The Dirac equation describing spin-1/2 particles is usually considered {\em in a
lagrangian framework} (in general, the canonical formalism is adopted), obtained from a
least action principle. After an
interesting preliminary study of the problem of the vibrating string, where Majorana
obtains a (classical) Dirac-like equation for a two-component field, he then passes to
consider a semiclassical relativistic theory for the electron, wherein the Klein-Gordon
equation and the Dirac equation are deduced from a semiclassical Hamilton-Jacobi
equation. Later on, the field equations and their properties (Lorentz invariance, issues
related to the probabilistic interpretation, and so on) are considered in detail, and the
quantization of the (free) Dirac field is discussed by means of the standard formalism
with the use of annihilation and creation operators. Then, the electromagnetic
interaction is introduced in the Dirac equation and the superposition of the Dirac and
Maxwell fields is studied in a {\it peculiar} way, obtaining the expression for the
quantized Hamiltonian of the interacting system after a normal mode decomposition.

\h Real (rather than complex) Dirac fields, introduced by Majorana in his famous paper
(9) on the symmetrical theory for electrons and positrons are considered in some points in
the Quaderni, and by two slightly different formalisms
(different decompositions of the field). The introduction of the electromagnetic
interaction is performed in a rather interesting fashion, and an {\it explicit} expression
for the total angular momentum carried by the real Dirac field is obtained starting from
the Hamiltonian.

\h  Some work behind Majorana's important paper (7) can be found
as well in the Quaderni. Here the
author, starting from the usual Dirac equation for a 4-component
spinor, obtains {\it explicit} expressions for the Dirac matrices
in the cases of 6-component and 16-component spinors. Interesting
enough, at the end of his discussion, Majorana also treats the
case of spinors with an {\it odd} number of components, namely a
5-component field.

\h  For what concerns quantum electrodynamics, again it is
generally considered in a lagrangian and hamiltonian framework,
with the use of a least action principle. As it is {\it now}
usual, the electromagnetic field is decomposed in plane wave
operators, and its properties are studied in a {\it full
Lorentz-invariant formalism} by employing group-theoretical
arguments. Explicit expressions for the quantized hamiltonian,
creation and annihilation operators for the photons, as well as
angular momentum operator, are deduced in several different basis,
along with the appropriate commutation relations. The attitude of Majorana
is near to that (more
mathematical and more efficient) introduced by Heisenberg, Born,
Jordan and Klein, rather than to the famous one by Fermi: once more
anticipating the times.

\h  In the {\em Quaderni} one also finds several studies, following
an original idea by Oppenheimer, aimed at exploring the
possibility of describing the electromagnetic field by means of the
Dirac formalism. Some
emphasis is given to consider the properties of the
electromagnetic field as described by a real wavefunction for the
photon\cite{MBR,Giannetto,SE}, a study which
is well beyond the contemporary works of other authors. Other
interesting investigations concern the possibility to introduce an
{\it intrinsic} time delay (as a universal constant) in the
expressions for the electromagnetic retarded fields, and a
study about the modification of the Maxwell
equations in the presence of magnetic monopoles.

\h  Besides the theoretical work about quantum
electrodynamics, some applications are as well considered. This is
the case of the free electron scattering, where Majorana gives an explicit expression
for the transition probability, and the coherent scattering of
bound electrons. Several other
scattering processes are as well considered in the framework of perturbation theory, by means of
the Dirac method or the Born method.

\h  As remarked above, the most known (to the contemporary physicists community)
contribution by Majorana in nuclear physics was his theory of nuclei made of protons and
neutrons interacting through an exchange force of a particular kind (which corrected the
Heisenberg model). In the research notes of the Quaderni several pages are also devoted to
study possible forms of the nucleon
potential inside a given nucleus, describing the interaction between neutrons and
protons. Although generic nuclei are often considered in the discussion, some particular
care is given by the author to light nuclei (deuteron, $\alpha$-particles, etc.): the studies performed by Majorana on
this subject were, at the same time, preliminary studies as well as and generalizations of what
published by himself in his well-known paper (8), revealing a very rich, and
peculiar, reasoning.

\h In addition to this, other nuclear physics topics were
considered by our author (which appeared in the {\em Volumetti} too),
and here we only mention the study of the problem of the
energy loss of $\beta$ particles in passing through a medium,
where he deduced the Thomson formula by using classical arguments.

\h The largest number of pages in the Quaderni is, however,
devoted to atomic physics, in
agreement with the fact that such a topic was the main research
argument faced by the Fermi group in Rome in the years 1928-1933.
This is also testified by the papers published by Majorana in
those years. Some echo of the work reported in those papers is
present in the Quaderni, that point out that, especially in
the case of the article (5) on the incomplete $P^\prime$ triplets,
some interesting material was not included by the author in
the published work.

\h Several expressions for the wavefunctions and the different
energy levels of two-electron atoms (and, in particular, of
helium) are considered by Majorana, mainly in the framework of a
variational method aimed at solving the related Schr\"odinger
equation. Numerical values for the corresponding energy terms are
usually reported by Majorana in large summary tables. Some approximate
expressions are also obtained for three-electron atoms (and, in
particular, for lithium) and for alkali, including the effect of
polarization forces in hydrogen-like atoms. \ The problem of the hyperfine structure of the energy spectra of
complex atoms is considered in some detail as well, revealing a
careful attention of Majorana to the existing literature. A
generalization of the Land\`e formula for the hyperfine splitting
to {\em non-Coulomb} atomic field is given, together with a {\em
relativistic} formula for the Rydberg corrections of the hyperfine
structures. Such a detailed study by Majorana forms the basis of
what discussed by Fermi and Segr\`e in a well-known paper of 1933
on this topic, as acknowledged by those authors themselves.

\h A smaller part of the Quaderni is devoted to several problems of molecular
physics. Majorana studied in some detail, for example, the
helium molecule and, then, considered the general theory of the vibrational modes in
molecules, with particular reference to the $C_2H_2$ molecule of acetylene (which
presents peculiar geometric properties).

\h  Important are also some other pages, where the author considered the problem of ferromagnetism in the framework
of the Heisenberg model with an exchange interaction. The approach used by Majorana in
this study, however, is original, since it does follow neither the Heisenberg
formulation, nor the subsequent van Vleck formulation in terms of the spin Hamiltonian. By
using statistical arguments, Majorana calculated the magnetization (with respect to the
saturation value) of the ferromagnetic system when an external magnetic field is applied
on it, and the spontaneous magnetization. Several examples of ferromagnetic materials,
with different geometries, were as well reported.

\h  A number of other interesting topics, even dealing with what
Majorana encountered in his academic studies at the University of
Rome, may be found as
well in the Quaderni. This is the case, for example, of the
problem of the electromagnetic and {\it electrostatic} mass of the
electron (a problem considered by Fermi in one of his
famous papers of 1924), or of his studies on tensor calculus,
following Levi-Civita. We do not go further in our
discussion, our aim being only that of drawing the
attention of the reader on some specific points.

\h Let us add a comment, however. The method followed by Majorana for composing his notes was
the following. When he reached a ``semi-definitive" result
in studying a given topic in his first-hand notes, he reported
such a result into a {\em Quaderno}. Subsequently, after further
research on the topic considered, if Majorana reached a
``conclusive" result on this (according to his opinion), he
reported this final result in a {\em Volumetto}. Clear expositions of particular
topics can, then, be found only in the {\em Volumetti}.

\h The 18 {\em Quaderni} deposited at the Domus Galilaeana are
booklets, composed of about 200
pages each, accounting for a total of about 2800 pages.
Differently from what happens for the {\em Volumetti}, in the
{\em Quaderni} no date is present, except for
the Quaderni no.16 (``1929-30''), no.17 (``started on 20 June
1932'') and, probably, no.7 (``about year 1928'').

\h A Bibliography follows. Far
from being exhaustive, it provides nevertheless some references
about the topics touched upon in this paper.

\

\section{Acknowledgments}
This work has been partially supported by FAPESP, Brazil (under
grant 2013/12025-8), and thanks are due to Michel Zamboni-Rached for his
collaboration, and to Hugo Hern\'andez-Figueroa for kind interest.  \ We are moreover
indebted to Dr. Carlo Segnini, the former
curator of the Domus Galileana at Pisa (as well as to previous
curators and directors), who facilitated our access to the original
manuscripts, and to Salvatore Esposito for continuous collaboration.  Last but not least, the author wishes
to thank Dharam V. Ahluwalia, and Cheng-Yang Lee, for kind invitation to
the `` 2nd International Workshop on Elko and Mass Dimension One Fermions " and subsequent quite useful editorial help.

\

\

{\em Observation: Most of the material contained in book \cite{ER} is covered by copyright
in favour of Maria Majorana, a late Ettore's sister, and the present author. \ It includes
letters and photos. \ It has been decided, however, that for scientific purposes the letter
excerpts appearing above may be reproduced provided that also the original source (ref.\cite{ER})
is duly quoted. \ The photos entering this paper cannot be further reproduced [incidentally, they
have been also {\em deposited} by the author c/o the ``Emilio Segr\'e Visual Archives" of the AIP's Niels
Bohr Library \& Archives] -- The author.}

\

\end{document}